%% file: main.tex
\documentclass[conference]{IEEEtran}
\IEEEoverridecommandlockouts
\usepackage{cite}
\usepackage{amsmath,amssymb,amsfonts}
\usepackage{algorithmic}
\usepackage{graphicx}
\usepackage{textcomp}
\usepackage{xcolor}

\usepackage{booktabs} 

\usepackage[english]{babel}
\usepackage{moresize}
\usepackage{amsmath}

\usepackage{algorithmic} 
\usepackage{balance}
\usepackage{comment}
\usepackage{paralist}
\usepackage{multirow}

\usepackage{filecontents}

\usepackage{flushend}
\usepackage[english]{babel}
\usepackage[latin1]{inputenc}
\usepackage{mathrsfs}
\usepackage{graphicx}
\usepackage{amssymb}
\usepackage{url}
\usepackage{subfigure}
\usepackage{amsmath}
\usepackage{enumitem}
\usepackage[linesnumbered,algoruled,boxed,lined]{algorithm2e}
\usepackage{adjustbox}
\usepackage{amssymb}
\usepackage{times}
\usepackage{hyperref}

\usepackage{pgfplots}
\usetikzlibrary{pgfplots.dateplot}

\usepackage{filecontents}
\definecolor{tblue}{RGB}{31,119,180}
\definecolor{torange}{RGB}{255,127,14}
\definecolor{tgreen}{RGB}{44,160,44}
\definecolor{tred}{RGB}{214,39,40}
\definecolor{tpurple}{RGB}{148,103,189}

\usepackage{filecontents}

\newcommand{\hide}[1]{} 

\newcommand{\etal}{\textit{et al}.}

\newcommand{\ie}{\textit{i}.\textit{e}.}
\newcommand{\eg}{\textit{e}.\textit{g}.} 
\newcommand{\wrt}{\textit{w}.\textit{r}.\textit{t}} 
\newtheorem{Dfn}{Definition}

\def\BibTeX{{\rm B\kern-.05em{\sc i\kern-.025em b}\kern-.08em
    T\kern-.1667em\lower.7ex\hbox{E}\kern-.125emX}}
    
\begin{document}


\title{Global Context Enhanced Social Recommendation with Hierarchical Graph Neural Networks

\thanks{\textbf{*Corresponding author: Yong Xu.}}}


\def\model{SR-HGNN}
\def\full{}



\author{\IEEEauthorblockN{Huance Xu$^\dag$, Chao Huang$^\S$, Yong Xu$^{\dag *}$, Lianghao Xia$^\dag$, Hao Xing$^\ddag$, Dawei Yin$^\natural$}
\IEEEauthorblockA{\textsuperscript{$\dag$}South China University of Technology, \textsuperscript{$\S$}JD Finance America Corporation, \textsuperscript{$\ddag$}VIPS Research, \textsuperscript{$\natural$}Baidu inc \\
\{cshuance.xu, yxu, cslianghao.xia\}@mail.scut.edu.cn,\\ chaohuang75@gmail.com, hao.xing@vipshop.com, yindawei@acm.org}}


\maketitle

\begin{abstract}
Social recommendation which aims to leverage social connections among users to enhance the recommendation performance. With the revival of deep learning techniques, many efforts have been devoted to developing various neural network-based social recommender systems, such as attention mechanisms and graph-based message passing frameworks. However, two important challenges have not been well addressed yet: (i) Most of existing social recommendation models fail to fully explore the multi-type user-item interactive behavior as well as the underlying cross-relational inter-dependencies. (ii) While the learned social state vector is able to model pair-wise user dependencies, it still has limited representation capacity in capturing the global social context across users. To tackle these limitations, we propose a new \underline{S}ocial \underline{R}ecommendation framework with \underline{H}ierarchical \underline{G}raph \underline{N}eural \underline{N}etworks (\model). In particular, we first design a relation-aware reconstructed graph neural network to inject the cross-type collaborative semantics into the recommendation framework. In addition, we further augment \model\ with a social relation encoder based on the mutual information learning paradigm between low-level user embeddings and high-level global representation, which endows \model\ with the capability of capturing the global social contextual signals. Empirical results on three public benchmarks demonstrate that \model\ significantly outperforms state-of-the-art recommendation methods. Source codes are available at: \emph{https://github.com/xhcdream/SR-HGNN}.
\end{abstract}


\input{intro}
\input{model}
\input{solution}

\input{eval}
\input{relate}

\input{conclusion}

\section*{Acknowledgments}
We thank the anonymous reviewers for their constructive feedback and comments. This work is supported by National Nature Science Foundation of China (61672241), Major Project of National Social Science Foundation of China (18ZDA062), Natural Science Foundation of Guangdong Province (2016A030308013), Science and Technology Program of Guangdong Province (2019A050510010).

\bibliographystyle{abbrv}
\bibliography{refs} 

\end{document}

%% file: intro.tex
\section{Introduction}
\label{sec:intro}

Recommender systems have play an important role in meeting user's personalized interests and alleviating the information
overload for various applications, ranging from e-commence platforms~\cite{niu2020dual,huang2019online}, content provider~\cite{wu2019npa,shi2020salience} to online review systems~\cite{tao2019multi,ntf2019neural}. With the prevalence of social networks in real-life online applications~\cite{song2016influential,2016topic}, a key line of research work seeks to boost the recommendation performance via exploiting the users' social relationships (\eg, online friends)~\cite{liu2019discrete}. In social recommender systems, users' social ties serve as the important side information to provide connectivity information and semantic relatedness between users, and thus are utilized to enhance transitional recommendation models in yielding better results~\cite{wu2018socialgcn,yu2019generating,socialrec2017}.

The core challenge of exploring social information in recommender systems is: how to incorporate user-user relationships into the collaborative filtering scenario with interactive pattern learning between users and items~\cite{wu2019neural,chen2019samwalker}. Conventional social-aware recommendation methods have made significant process to regularize the matrix factorization framework with social information of users~\cite{ma2011recommender,jamali2010matrix,yang2016social}. Recently, the immense success of deep learning techniques has witnessed some research work on the exploration of neural network structures to enhance recommender systems with social signals. Specifically, there are several attempts which adopt attentive memory network for attending to certain parts when performing relation encoding among users~\cite{chen2019social,chen2019efficient}. For example, Chen~\etal~\cite{chen2019efficient} introduced a transfer neural network to model the interplay between the social and item interaction domain. Additionally, the social influence among users have been approximated with a neural diffusion scheme through the layer-wise propagation of user embeddings~\cite{wu2019neural}. In view of recent advancements of graph neural networks, a handful of graph-based message passing structures have been developed to aggregate relation structural information over the constructed graph with users and items~\cite{fan2019graph,yu2020enhance}.


Despite the effectiveness of the above solutions, two important challenges have not been well addressed yet. \emph{First}, practical recommendation scenarios may involve different types of user-item interactive behaviors, such as users' different ratings over items in online review systems, or different activities of customers (\eg, browse, purchase) in e-commerce sites~\cite{gao2019neural,xia2020multiplex}. However, most of existing social recommender systems either ignore the multi-type nature of user-item interactions, or assume that different types of relation edges between users and items share the same representation space (\eg, learning with categorical one-hot encoding~\cite{chen2019social,chen2019efficient} or continuous value-based linear transformation~\cite{liu2018social,wang2019social}). Such relation heterogeneity could provide auxiliary behavior semantics that can hardly be comprehensively captured by current social-aware recommendation models. While intuitively useful to integrate multi-type user-item interactions into the learning of user preference, it is non-trivial to deal with it well. In particular, the complex dependencies across different interactions, making it difficult to distill the desired relation-aware collaborative signals with the joint incorporation of high-order connectivity from both user and item dimensions.

\emph{Second}, the current designed embedding functions of users' social information, lack an effective encoding of high-order relational structures, which is latent in user-user relations to reveal the social similarity across users~\cite{yang2017fast}. To be more specific, most of current methods fuse cross-user relations from direct neighbourhood~\cite{chen2019social,chen2019efficient} and can hardly capture the high-order social influence between users. While GraphRec~\cite{fan2019graph} proposes to design graph structure-based neural network to aggregate relations between connected users, it only embeds the social signals into latent representation space with local proximity, due to the heavy computational cost in performing higher-order message passing over the social graph of users~\cite{abu2019mixhop}. Hence, how to jointly capture the local and global contextual signals of users' relationships in the recommendation framework remains a significant challenge. \\\vspace{-0.1in}

\noindent \textbf{Present Work}. In light of aforementioned challenges, we propose a new \underline{\textbf{S}}ocial \underline{\textbf{R}}ecommendation model with \underline{\textbf{H}}ierarchical \underline{\textbf{G}}raph \underline{\textbf{N}}eural \underline{\textbf{N}}etwork (\model). In the two-phase recommendation framework, we propose to handle user-item interaction heterogeneity through a relation-aware reconstructed graph neural module. This graph neural network architecture automatically extracts the multi-relation collaborative signals from user-item interactions. We further supercharge the relation-aware graph learning framework with the global information to reconstruct the cross-domain (user-user and user-item) relations, during the embedding process of the graph neural network as constraints.

Additionally, \model\ captures users' social relations by advancing the graph-based neural relation encoder, to jointly capture local and global relational structures between users. The global context enhanced social encoder not only learns the low-level patch embeddings of users from their social neighbors, but also derives the high-level contextual signals of the social graph to augment the user representation process under a global graph-structured mutual information maximization architecture. Our \model\ generalizes the paradigm of mutual information estimation~\cite{hjelm2018learning,velivckovic2018deep} from feature vector space to the social graph-based user relation modeling, which injects hierarchical social similarities into relation learning via discriminating corrupted social structures.


The contributions of this paper are highlighted as follows:

\begin{itemize}[leftmargin=*]
\item We highlight the critical importance of preserving both global structure of social dependencies and multi-typed interactive patterns between users and items in social recommendation task. Towards this goal, we propose \model, a new social recommender system with hierarchically structured graph neural networks.\\\vspace{-0.1in}

\item In \model\ framework, we propose a relation-aware reconstructed graph neural module to i) encode the collaborative signal in the form of multi-type user-item relations, and ii) inject the social-aware multi-relational information into the embedding process via reconstructing the global connectivities between users and items.\\\vspace{-0.1in}

\item To fully explore the global structural contexts of social connections, we further propose to capture the users' dependencies with graph-level mutual information maximization. This designed social relation encoder uniforms feature representation spaces from both low-level (locally) to high-level (globally) social contextual signals.\\\vspace{-0.1in}

\item Experimental results on three real-world datasets show the superiority of our \model\ framework over various baselines in yeilding better reccommendation results.
\end{itemize}

%% file: model.tex
\section{Preliminaries and Problem Definition}
\vspace{-0.05in}
\label{sec:model}

We begin with some necessary notations and then formally present the our studied social recommendation problem. We consider a recommendation scenario where a group of $M$ users $U=\{u_1,...,u_m,...,u_M\}$ and a set of $N$ items $V=\{v_1,...,v_n,...,v_N\}$. We further define relevant inputs as below:\\\vspace{-0.1in}

\begin{Dfn}
\textbf{User Social Graph} $G_s$. We define the user social graph $G_s=\{U,E_s\}$ to represent the users' social relationships, where $U$ ($u_m\in U$) and $E_s$ denotes the set of user nodes and edges between them. In specific, if two users $u_m$ and $u_{m'}$ are socially connected, there exists an edge between $u_m$ and $u_{m'}$ in the constructed user social graph $G_s$.\\\vspace{-0.1in}
\end{Dfn}

\begin{Dfn}
\textbf{Multi-Type Interaction Graph} $G_r$. With the consideration of different interactions between users and items, a multi-typed interaction graph is defined as $G_r=\{U,V,E_r\}$, where $V$ represents the set of item nodes. Furthermore, $E_r$ denotes multiple types of interactive relations (\eg, different ratings or activities) between user $u_m$ and item $v_n$. \\\vspace{-0.12in}
\end{Dfn}

\noindent \textbf{Problem Statement}. Based on the aforementioned definitions, the studied social recommendation problem is formally defined as follows: \textbf{Input}: the user-user social relation data represented with user social graph $G_s$ and the user-item interaction data exhibited with the multi-type interaction graph $G_r$. \textbf{Output}: A predictive function which aims to estimate the unknown multi-typed user-item interactive relations in graph $G_r$.

%% file: solution.tex
\section{Methodology}
\label{sec:solution}

In this section, we elaborate the technical details of our developed \model\ framework. We first describe our relation-aware reconstructed graph neural network to capture the multi-typed user-item interactions. Then, we present the designed social dependency encoding framework which contextualizes the relation-aware collaborative signal modeling architecture with the global social context.

\subsection{Multi-Typed User-Item Interactive Relation Learning}
\begin{figure}
	\centering
	\includegraphics[width=0.48\textwidth]{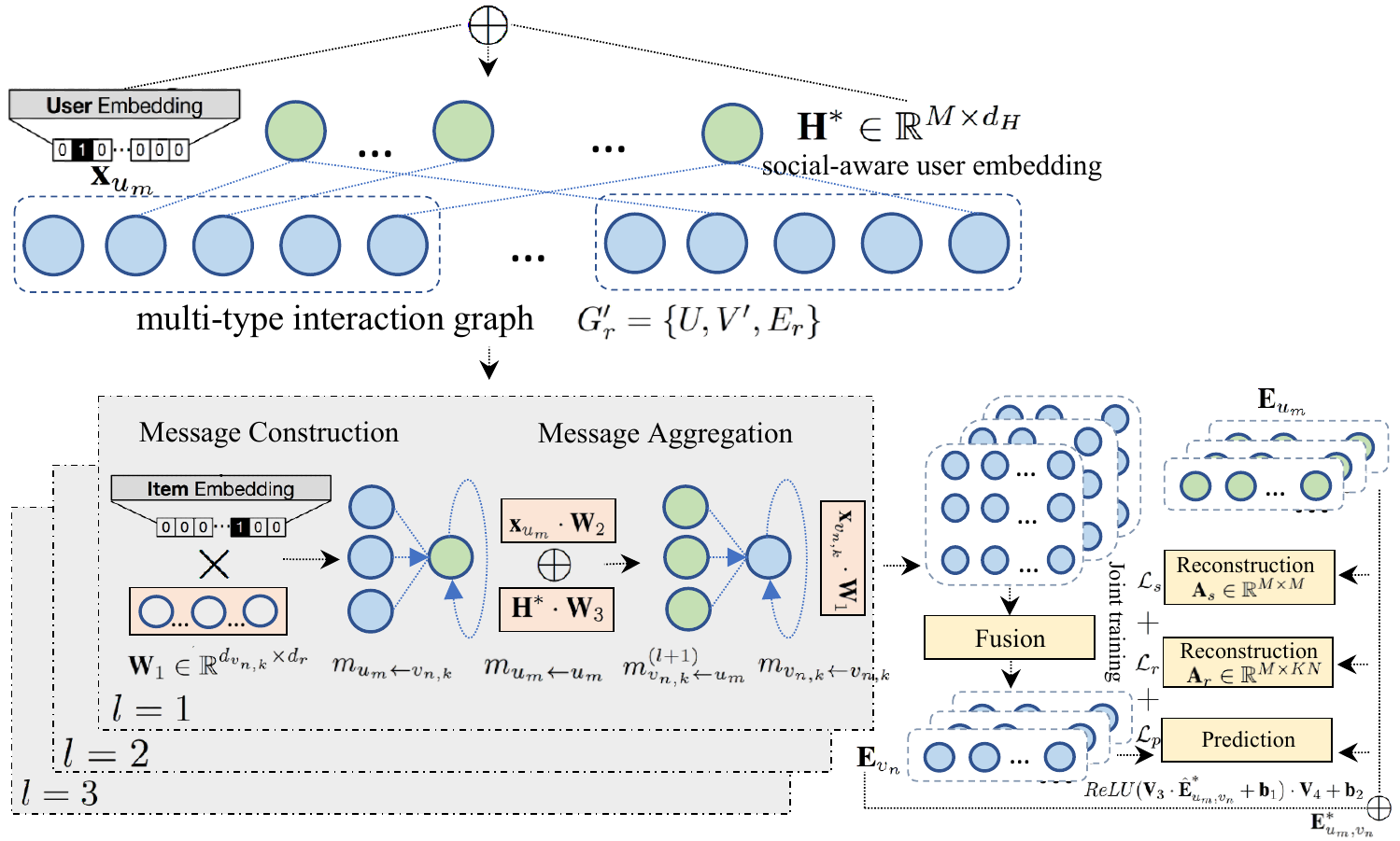}
	\vspace{-0.1in}
	\caption{The model architecture of multi-typed user-item interactive relation learning in \model\ framework.}
	\label{fig:framework_2}
	\vspace{-0.1in}
\end{figure}

To learn the multi-interactive collaborative signals, we develop a relation-aware message-passing architecture (as shown in Figure~\ref{fig:framework_2}) between users and items with the differentiation of different user-item interactive relations. In particular, we first decompose the item vertex of multi-type interaction graph $G_r$ into multiple sub-nodes: $v_n\rightarrow (v_{n,1},...,v_{n,K})$, where $K$ is the number of interaction types. Each sub-node $v_{n,k}$ is connected to the corresponding user $u_m$ with the $k$-th type of user-item relation. By doing so, the multi-typed relations between user and item are reflected on the updated graph $G_r'=\{U,V',E_r\}$ with the total number of $(M+N\times K)$ vertices.

\subsubsection{\bf Embedding Propagation Module}
To encode collaborative similarity across users and items, we design a message-passing graph neural network to leverage the multi-relation user-item interaction graph $G_r'$ for embedding propagation. In general, the message passing architecture consists of two key components: \emph{message construction} and \emph{message aggregation}.

\textbf{Message Construction Phase}. We define our message passing from user $u_m\in U$ to his interacted relation-specific item $v_{n,k}\in V'$ as follows:
\begin{align}
\label{eq:message_passing}
m_{u_m \leftarrow v_{n,k}} = f(\textbf{x}_{v_{n,k}}, \lambda_{m,n}^k)
\end{align}
\noindent where $f(\cdot)$ is the message encoding function. $\textbf{x}_{v_{n,k}}$ is the input feature representation corresponds to $n$-th item node interacted with $u_m$ given the $k$-th interaction type. $\lambda_{m,n}^k$ denotes the decay factor for the propagation between $u_m$ and $v_{n,k}$. In our \model, we define encoding function $f(\cdot)$ as below:
\begin{align}
m_{u_m \leftarrow v_{n,k}} = \frac{1}{\sqrt{|J_m||J_{n,k}|}} (\textbf{x}_{v_{n,k}} \cdot \textbf{W}_1)
\end{align}
\noindent where $J_m$ represents the set of item sub-nodes interacted with user $u_m$, and $J_{n,k}$ denotes the set of users that are connected with $v_{n,k}$. $\textbf{W}_1 \in \mathbb{R}^{d_{v_{n,k}}\times d_r}$ is the weight matrix, where $d_{v_{n,k}}$ and $d_r$ represents the latent dimensionality of $\textbf{x}_{v_{n,k}}$ and the propagation module, respectively. Based on the convolutional operation, $\lambda_{m,n}^k$ reflects that the influence strength of $v_{n,k}$ over $u_m$ is inversely proportional to the number of $u_m$'s connected nodes, \ie, $\lambda_{m,n}^k=\frac{1}{\sqrt{|J_m||J_{n,k}|}}$.

Similarly, we define the message encoding function from user $u_m$ to item sub-node $v_{n,k}$ as follows:
\begin{align}
m_{v_{n,k} \leftarrow u_m} = \frac{1}{\sqrt{|J_m||J_{n,k}|}} (\textbf{x}_{u_m} \cdot \textbf{W}_2 \oplus \textbf{H}^* \cdot \textbf{W}_3)
\end{align}
\noindent $\textbf{H}^* \in \mathbb{R}^{M\times d_H}$ denotes the social-aware user representations which are learned from our designed social dependency encoding framework (as elaborated in Section~\ref{sec:social}). $\textbf{W}_2 \in \mathbb{R}^{d_{u_m} \times \frac{d}{2}}$ and $\textbf{W}_3 \in \mathbb{R}^{d_H \times \frac{d}{2}}$ are trainable transformation matrices.

\noindent \textbf{Message Aggregation Phase}. After obtaining the information from interacted users/items, we define our message aggregation function as follows:
\begin{align}
\textbf{E}_{u_m} &= \delta \Big (m_{u_m \leftarrow u_m} + \sum_{(n,k)\in J_m} m_{u_m \leftarrow v_{n,k}} \Big) \nonumber\\
\textbf{E}_{v_{n,k}} &= \delta \Big (m_{v_{n,k} \leftarrow v_{n,k}} + \sum_{u_m\in J_{n,k}} m_{v_{n,k} \leftarrow u_m} \Big)
\end{align}
\noindent $\delta(\cdot)$ is defined as the PReLU activation function. $m_{u_m \leftarrow u_m}$ and $m_{v_{n,k} \leftarrow v_{n,k}}$ respectively represents the self-propagated information for $u_m$ and $v_{n,k}$ with the formal definitions:
\begin{align}
m_{u_m \leftarrow u_m} & = \frac{1}{\sqrt{|J_m|}} (\textbf{x}_{u_m} \cdot \textbf{W}_2 \oplus \textbf{H}^* \cdot \textbf{W}_3)\nonumber\\
m_{v_{n,k} \leftarrow v_{n,k}} & = \frac{1}{\sqrt{|J_{n,k}|}} (\textbf{x}_{v_{n,k}} \cdot \textbf{W}_1)
\end{align}

Based on the aforementioned message passing and aggregation functions, we will present how to incorporate high-order relationships across users and items, into our multi-typed interactive relation learning framework. We formally define our high-order propagation process as:
\begin{align}
m_{u_m \leftarrow v_{n,k}}^{(l)} &= \lambda_{m,n}^k (\textbf{E}_{v_{n,k}}^{(l-1)} \textbf{W}_1^{(l)}) \nonumber\\
m_{u_m \leftarrow u_m}^{(l)} &= \frac{1}{\sqrt{|J_m|}} (\textbf{E}_{u_m}^{(l-1)} \textbf{W}_2^{(l)}) \nonumber\\
m_{v_{n,k} \leftarrow u_m}^{(l)} &= \lambda_{m,n}^k (\textbf{E}_{u_m}^{(l-1)} \textbf{W}_2^{(l)}) \nonumber\\
m_{v_{n,k} \leftarrow v_{n,k}}^{(l)} &= \frac{1}{\sqrt{|J_{n,k}|}} (\textbf{E}_{v_{n,k}}^{(l-1)} \textbf{W}_1^{(l)}) \nonumber
\end{align}
\noindent $l$ is the index of graph neural network layers $L$. The propagation can be rewrote with the matrix form based on the $(l-1)$-th order node representation $\textbf{E}^{(l-1)}$ and weight matrix $\textbf{W}^{(l)}$ from the $l$-th layer as follows:
\begin{align}
\textbf{E}^{(l-1)} \textbf{W}^{(l)} \in \mathbb{R}^{(M+NK)\times d} = 
\begin{bmatrix}
\textbf{E}_{u_m}^{(l-1)} \cdot \textbf{W}_2^{(l)} \\ \textbf{E}_{v_{n,k}}^{(l-1)} \cdot \textbf{W}_1^{(l)}
\end{bmatrix}
\end{align}
\noindent We generate the final embeddings of user $u_m$ and item sub-node $v_{n,k}$ with the following concatenate operation:
\begin{align}
\label{eq:highOrder_embed}
\textbf{E}_{u_m} &= (\textbf{E}_{u_m}^{(1)} \oplus \textbf{E}_{u_m}^{(2)} \oplus \cdots \oplus \textbf{E}_{u_m}^{(L)})\nonumber\\
\textbf{E}_{v_{n,k}} &= (\textbf{E}_{v_{n,k}}^{(1)} \oplus \textbf{E}_{v_{n,k}}^{(2)} \oplus \cdots \oplus \textbf{E}_{v_{n,k}}^{(L)})
\end{align}
\noindent the overall item representation $\textbf{E}_{v_{n}}$ is aggregated over the set of $\{\textbf{E}_{v_{n,1}},...,\textbf{E}_{v_{n,K}}\}$ with the mean pooling operation.



\subsubsection{\bf Reconstruction-based Context Incorporation}
\label{sec:reconstruction}
To inject the social-aware cross-relational signals into our multi-typed user-item interaction encoding architecture, we augment our graph-based message passing module with the exploration of both global social (user-user) and multi-interactive (user-item) contexts. In specific, we incorporate the cross-domain reconstruction constrains (\ie, adjacent matrices $\textbf{A}_s \in \mathbb{R}^{M\times M}$ of graph $G_s$ and $\textbf{A}_r \in \mathbb{R}^{M\times KN}$ of graph $G_r$) into embedding process of $\textbf{E}_{u_m}$ and $\textbf{E}_{v_{n,k}}$. We adopt the pairwise 
BPR loss as the objective in the reconstruction phase. For the reconstruction of user-item interactions $\textbf{A}_r$: if user $u_m$ interacts with item $v_n$ with the $k$-th behavior relation, we will sample the corresponding negative samples $v_{n,k^-}$ from other non-interacted $(K-1)$ relations. For the reconstruction of user-user social relations $\textbf{A}_s$, the negative instance $u_{m^-}$ is sampled from his non-connected users. Formally, we present the reconstruction-based context incorporation as follows:
\begin{align}
\label{eq:recon_loss}
s_{m,n,k}^{A_r} &= \delta((\textbf{E}_{u_m} \oplus \textbf{E}_{v_{n,k}}) \textbf{V}_1 + \textbf{b}_r) \textbf{W}_4 \nonumber\\
\mathcal{L}_{r} &= -\frac{1}{\psi(\textbf{A}_r)} \sum_{(m,n,k^+) \in O_r} \log \sigma(s_{m,n,k^+}^{A_r} - s_{m,n,k^-}^{A_r}) \nonumber\\
s_{m,m'}^{A_s} &= \delta((\textbf{E}_{u_m} \oplus \textbf{E}_{u_{m'}}) \textbf{V}_2 + \textbf{b}_s) \textbf{W}_5 \nonumber\\
\mathcal{L}_{s} &= -\frac{1}{\psi(\textbf{A}_s)} \sum_{(m,m^+) \in O_s} \log \sigma(s_{m,m^+}^{A_s} - s_{m,m^-}^{A_s})
\end{align}
\noindent where $\textbf{V}_1$, $\textbf{V}_2$, $\textbf{W}_4$ and $\textbf{W}_5$ are learnable weight matrices. $\textbf{b}_r$, $\textbf{b}_s$ are bias terms. $\psi(\textbf{A}_r)$, $\psi(\textbf{A}_s)$ indicates the number of non-zero elements in $\textbf{A}_r$ and $\textbf{A}_s$, respectively.

\subsection{Global Context Enhanced Social Dependency Modeling}
\label{sec:social}

\begin{figure}
	\centering
	\includegraphics[width=0.47\textwidth]{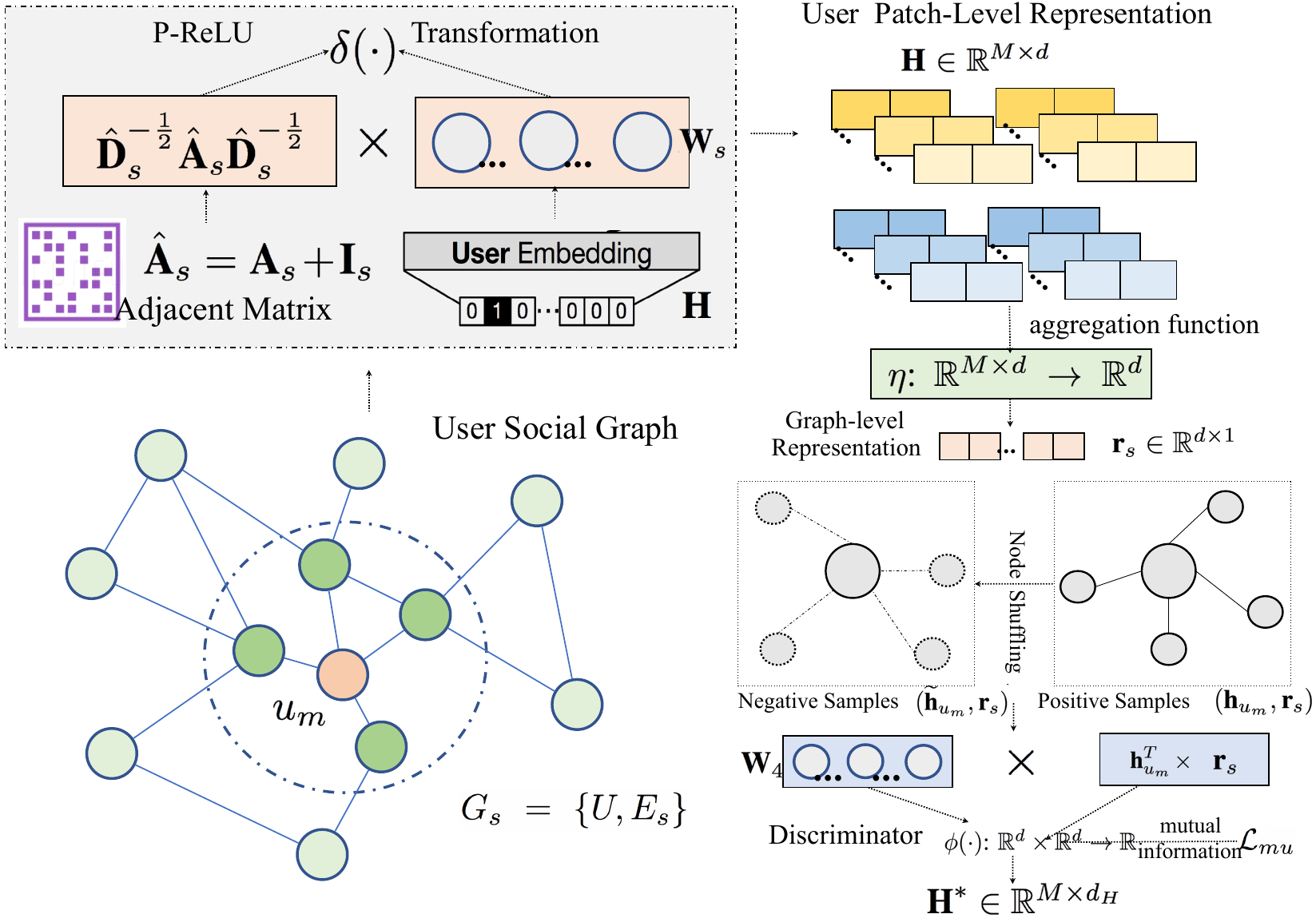}
	\caption{The model architecture of the global social dependency encoder.}
	\label{fig:framework_1}
\end{figure}

To jointly capture the local and global social dependencies, we further develop a mutual information-based graph learning module to distill the hierarchical social similarity in the user embedding space. We build our social relation encoding module upon a dual-stage graph learning architecture (as shown in model architecture Figure~\ref{fig:framework_1}).

We first design our graph-structured message propagation layer, to generate node-level latent representation $\textbf{h}_{u_m}\in \mathbb{R}^{d}$ of each individual user $u_m$, where $d$ indicates the hidden state dimensionality. We define the local information encoding function over the user social graph $G_s$ as follows:
\begin{align}
\label{eq:patch_embed}
\textbf{H} = \delta(\textbf{A}_s, \textbf{H} \textbf{W}^l_s)=\delta(\hat{\textbf{D}}^{-\frac{1}{2}}_s \hat{\textbf{A}}_s \hat{\textbf{D}}^{-\frac{1}{2}}_s \textbf{H} \textbf{W}_s)
\end{align}
\noindent where $\delta(\cdot)$ denotes the non-linear activation function Parametric ReLU~\cite{he2015delving} and $\textbf{H} \in \mathbb{R}^{M\times d}$ corresponds to the encoded representations of all users. To inject the self-propagated signals, the identity matrix $\textbf{I}_s$ is added into the adjacent matrix $\textbf{A}_s$ (constructed from graph $G_s$) to generate $\hat{\textbf{A}}_s$, $\hat{\textbf{A}}_s=\textbf{A}_s+\textbf{I}_s$. A symmetric normalization strategy is applied in performing the neighboring information aggregation with the operation of $\hat{\textbf{D}}^{-\frac{1}{2}}_s \hat{\textbf{A}}_s \hat{\textbf{D}}^{-\frac{1}{2}}_s$, where $\hat{\textbf{D}}_s$ represents the diagonal node degree matrix of $\hat{\textbf{A}}_s$.


After learning the node-level user embeddings $\textbf{H} \in \mathbb{R}^{M\times d}$ encoded from social structured graph, our next step is to obtain the graph-level representation over the social graph $G_s$. We first define graph-level aggregation function $\eta$: $\mathbb{R}^{M\times d} \rightarrow \mathbb{R}^d$ with the consideration of node degrees as follows:
\begin{align}
\label{eq:graph_rep}
\textbf{r}_s= \sigma \Big ( \frac{\sum_{m=1}^M \textbf{H}_m \cdot b_{m,m}}{\sum_{m=1}^M \sum_{m'=1}^M a_{m,m'}} \Big )
\end{align}
\noindent where $\textbf{r}_s \in \mathbb{R}^{d\times 1}$ indicates the fused global latent representation of graph $G_s$. $\sigma$ denotes the sigmoid activation function. Furthermore, $b_{m,m}$ and $a_{m,m'}$ represents the element in the degree matrix $\hat{\textbf{D}}_s$ and adjacent matrix $\hat{\textbf{A}}_s$, respectively.


Inspired by the paradigm of mutual information maximization in feature representation~\cite{velivckovic2018deep}, we enhance our social relation embeddings with the exploration of mutual information between node-level user embedding $\textbf{H}$ and graph-level representation $\textbf{r}_s$.
To encode the mutual relations in social graph $G_s$ and follow this paradigm, we propose to train a discriminator to differentiate positive samples and negative samples from social graph $G_s$ with the preservation of connected topological structure. Specifically, positive samples are denoted as $(\textbf{h}_{u_m}, \textbf{r}_s)$, and negative instances $(\widetilde{\textbf{h}}_{u_m}, \textbf{r}_s)$ are generated following the node shuffling strategy to associate each user with fake feature vectors $\textbf{H}^0$ with one-hot encoding. Then, we feed the generated positive $(\textbf{h}_{u_m}, \textbf{r}_s)$ and negative instances $(\widetilde{\textbf{h}}_{u_m}, \textbf{r}_s)$ into our defined discriminator function $\phi(\cdot)$: $\mathbb{R}^d \times \mathbb{R}^d \rightarrow \mathbb{R}$.
\begin{align}
\phi(\textbf{h}_{u_m}, \textbf{r}_s) = \sigma(\textbf{h}_{u_m}^T \cdot \textbf{W}_6 \cdot \textbf{r}_s)
\end{align}
\noindent where discriminator function $\phi(\cdot)$ aims to generate a probability score of user $u_m$ belongs to graph $G_s$ given the corresponding representations $(\textbf{h}_{u_m}, \textbf{r}_s)$. $\textbf{W}_6 \in \mathbb{R}^{d\times d}$ is the learnable transformation matrix. We further define our mutual information-based loss as follows:
\begin{align}
\label{eq:dgi_loss}
\mathcal{L}_{mu} &= - \frac{1}{N_{pos}+N_{neg}} \Big ( \sum_{i=1}^{N_{pos}} \rho(\textbf{h}_{u_m}, \textbf{r}_s) \cdot log \phi(\textbf{h}_{u_m}, \textbf{r}_s) \nonumber\\
&+ \sum_{i=1}^{N_{neg}} \rho(\widetilde{\textbf{h}}_{u_m}, \textbf{r}_s) \cdot log [1-\phi(\widetilde{\textbf{h}}_{u_m}, \textbf{r}_s)] \Big )
\end{align}
\noindent where $N_{pos}$ and $N_{neg}$ denotes the number of positive and negative samples, respectively. $\rho(\cdot)$ is an indicator function where $\rho(\textbf{h}_{u_m}, \textbf{r}_s)=1$ and $\rho
(\widetilde{\textbf{h}}_{u_m}, \textbf{r}_s)=1$ corresponds to positive and negative instance during the training phase. By minimizing the loss $\mathcal{L}_{mu}$ (maximizing the mutual information between local node-level and global graph-level representations), we could generate the enhanced user representations $\textbf{H}^* \in \mathbb{R}^{M\times d_H}$ with the preservation of global social context.

\subsection{Learning Process of \model}

\begin{algorithm}[t]
    \small
	\caption{Learning Process of \model}
	\label{alg:learn_alg}
	\LinesNumbered
	\KwIn{multi-typed user-item interaction graph $G_r$, user-user social graph $G_s$
	user-item interaction tensor, sample number $s$, maximum epoch number $E_1, E_2$, loss weights $\omega_1, \omega_2, \omega_r$, learning rate $\eta$}
	\KwOut{trained parameters in $\Theta$}
	Initialize all parameters in ${\Theta}$\\
    \For{$e=1$ to $E_1$}{
        Calculate the graph-level representation of $G_s$ according to Eq~\ref{eq:patch_embed} to Eq~\ref{eq:graph_rep}\\
        Calculate the mutual information-based loss $\mathcal{L}_{mu}$ according to Eq~\ref{eq:dgi_loss}\\
        \For{$\theta$ in the social dependency modeling module}{
            $\theta=\theta-\eta\cdot\partial\mathcal{L}_{mu}/\partial\theta$
        }
    }
    \For{$e=1$ to $E_2$}{
        Calculate the high-order embeddings $\textbf{E}_{u_m}, \textbf{E}_{v_{n,k}}$ according to Eq~\ref{eq:message_passing} to Eq~\ref{eq:highOrder_embed}\\
        
        Draw a mini-batch of $(s_{m,n,k^+}^{A_r}, s_{m,n,k^-}^{A_r})$ and $(s_{m,m^+}^{A_s}, s_{m,m^-}^{A_s})$ for reconstruction\\
        Calculate the reconstruction loss $\mathcal{L}_r, \mathcal{L}_s$ according to Eq~\ref{eq:recon_loss}\\
        Calculate the prediction loss $\mathcal{L}_p$ according to Eq~\ref{eq:pred_loss}\\
        $\mathcal{L}=\mathcal{L}_p+\omega_1\mathcal{L}_r+\omega_2\mathcal{L}_s+\omega_r\|\Theta\|_F^2$\\
        \For{$\theta$ in the interaction modeling graph neural network}{
            $\theta=\theta-\eta\cdot\partial\mathcal{L}/\partial\theta$
        }
    }
    return all parameters $\mathbf{\Theta}$
\end{algorithm}

In the prediction layer, we incorporate the learned latent representations of user and item ($\textbf{E}_{u_m}$, $\textbf{E}_{v_n}$) into Multilayer Perceptron module, which is formally represented as follows:
\begin{align}
\textbf{E}_{u_m,v_n}^* &= \textbf{E}_{u_m} \oplus \textbf{E}_{v_n} \nonumber \\
\hat{r}_{m,n} &= ReLU(\textbf{V}_3 \cdot \textbf{E}_{u_m,v_n}^* + \textbf{b}_1) \cdot \textbf{V}_4 + \textbf{b}_2
\end{align}
\noindent where $\textbf{V}_3 \in \mathbb{R}^{2d\times d}$, $\textbf{V}_4 \in \mathbb{R}^{d\times 1}$ are learned transformation matrices, and $\textbf{b}_1$, $\textbf{b}_2$ are bias terms. We define the loss in our prediction layer as:
\begin{align}
\label{eq:pred_loss}
\mathcal{L}_p = \frac{1}{2} \sum_{m=1}^M \sum_{n=1}^N I_{m,n} (r_{m,n} - \hat{r}_{m,n})^2
\end{align}
\noindent where $I_{m,n}$ is the indicator function, \ie, $I_{m,n}=1$ if user $u_m$ is interacted with item $v_n$ and $I_{m,n}=0$ otherwise. $\hat{r}_{m,n}$ represents the explicit feedback between $u_m$ and $v_n$ corresponding to the different types of user-item relations. After incorporating the reconstruction factors stated in Section~\ref{sec:reconstruction}, we define our joint loss function as below:
\begin{align}
\label{eq:tot_loss}
\mathcal{L} = \mathcal{L}_p + \omega_{1} \mathcal{L}_r + \omega_{2} \mathcal{L}_s + \omega_{r} \left \| \Theta \right \|^2_F
\end{align}
\noindent $\omega_{1}$ and $\omega_{2}$ are parameters to the losses from different modules and prevent the overfitting issue. $\omega_{r}$ and $\Theta$ denotes the regularization term and model parameters, respectively. The training process is elaborated in Algorithm~\ref{alg:learn_alg}.

\subsection{Complexity Analysis of \model}
Next, we analyze the complexity of the proposed \model\ model. The embedding propagation module that learns graph-based representations costs $O((M+KN)\times d^2)$ computations for the message construction phase, and $O(N\times K\times M\times d)$ calculations for the message aggregation phase. By taking advantage of the sparse matrix-multiplication, the cost of the second phase is reduced to $O(\psi(\textbf{A}_r)\times d)$, where $\psi(\textbf{A}_r)$ denotes the number of user-item interactions. For the purpose of dimensionality reduction, $(M+KN)\times d$ is typically smaller than $\psi(\textbf{A}_r)$. Hence, the two phases cost $O(\psi(\textbf{A}_r)\times d)$. 

The reconstruction-based context incorporation utilizes $O(d^2)$ operations for each user-item pair, so $O(\psi(\textbf{A}_r)\times d^2)$ is required for the reconstruction. Analogously, we can find out the complexity of reconstructing the user-user interaction is $O(\psi(\textbf{A}_s)\times d^2)$. Totally, the complexity of the multi-typed user-item interactive relation learning is $O((\psi(\textbf{A}_r)+\psi(\textbf{A}_s))\times d^2)$, which is close to the common graph neural networks considering small $d$. The mutual information learning based social modeling propagates information in a similar way, and also costs $O(\psi(\textbf{A}_s)\times d)$ complexity for calculation. The efficiency of our \model\ is validated in the experiments by comparing the running time of our model with several state-of-the-arts.

%% file: eval.tex
\section{Evaluation}
\label{sec:eval}

To evaluate our \emph{\model}, we perform extensive experiments with three real-world recommendation datasets. Particularly, we aim to answer the following research questions:

\begin{itemize}[leftmargin=*]

\item \textbf{RQ1}: How is the performance of \emph{\model} when competing with various state-of-the-art recommendation methods?\\\vspace{-0.1in}

\item \textbf{RQ2}: What kind of benefit the developed key components in \emph{\model} (\eg, mutual information-based social relation encoder and relation-aware reconstructed graph neural module) can bring for social recommendation?\\\vspace{-0.1in}

\item \textbf{RQ3}: How does \textit{\model} perform \wrt\ different interaction sparsity levels as compared to competitive methods?\\\vspace{-0.1in}

\item \textbf{RQ4}: How do different hyperparameter settings affect the recommendation performance of our \emph{\model} model?\\\vspace{-0.1in}


\item \textbf{RQ5}: How is the model scalability of \emph{\model}?

\end{itemize}

\subsection{Experimental Settings}

\subsubsection{\bf Data Description}
We conduct performance validation with three real-world datasets: Epinions, Ciao and Douban. Table~\ref{tab:data} summarizes the statistics of these three datasets.\\\vspace{-0.15in}

\noindent \textbf{Epinions and Ciao}. Epinions and Ciao data is collected from the popular social networking-based consumer review site Epinions~\cite{fan2019graph} and Ciao~\cite{fan2019deep}, respectively. In these sites, users can establish social ties (who-trust-whom) with others, and interact with different items based on different rating scores (ranging from 1 to 5 and 1 as increment). We regard each rating score as an individual type of user-item interaction.\\\vspace{-0.1in}

\noindent \textbf{Douban}. This data is collected from the most popular Chinese online review platform: Douban. It is also a social networking platform which allows users to create connection with others based on their common interest. The rating interactions share the same score scales with Epinions and Ciao data.

\begin{table}[h!]
\vspace{-0.1 in}
\centering
\footnotesize
\caption{Statistics of Experimented Datasets.}
\vspace{-0.1 in}
\begin{tabular}{l| c| c| c}
\hline
Dataset & Ciao & Epinions & Douban\\
\hline
\# of Users & 7,375 & 22,164 & 50,694 \\
\# of Items & 105,060 & 296,277 & 24,088\\
\# of User-Item Interactions & 282,163 & 912,441 & 3,523,157\\
Interaction Density Degree & 0.0364\% & 0.0139\% & 0.2885\%\\
\# of Social Ties & 111,781 & 355,631 & 439,893\\
Social Tie Density Degree & 0.2055\% & 0.0724\% & 0.0171\%\\
\hline
\end{tabular}
\label{tab:data}
\end{table}

\subsubsection{\bf Evaluation Protocols}
We adopt two representative \emph{Root Mean Square Error (RMSE)} and \emph{Mean Absolute Error (MAE)} which have been widely used for recommendation with explicit feedback. Note that lower RMSE and MAE scores indicate better performance. Remarkable recommendation quality improvement can be achieved with small improvement on RMSE and MAE values~\cite{koren2008factorization,liu2018social}. Following the same settings in~\cite{fan2019graph}, we set the data percentage for training, validation and test set with $x\%$, $(1-x\%)/2$, $(1-x\%)/2$, respectively, where validation set is used for hyperparameter tuning. In our experiments, we set $x\%$ as ($60\%, 80\%$) to investigate the model performance with different input data ratio of user-item interactions.

\subsubsection{\bf Methods for Comparison}
We compare the \emph{\model} with state-of-the-art methods from different research lines:\\\vspace{-0.12in}

\noindent \textbf{Probabilistic Matrix Factorization Method}. We first consider the representative matrix factorization-based method.
\begin{itemize}[leftmargin=*]
\item \textbf{PMF}~\cite{mnih2008probabilistic}: it is a matrix factorization based probabilistic model which learns latent user/item feature vectors given their zero-mean spherical Gaussian priors.
\end{itemize}

\noindent \textbf{Conventional Social Recommendation Techniques}. We include several conventional social recommendation approaches which unify the user-item interaction and social relationships. 
\begin{itemize}[leftmargin=*]
\item \textbf{SocialMF}~\cite{jamali2010matrix}: this method incorporates the trust propagation in the matrix factorization architecture, to capture the social phenomenon in the recommendation scenario.
\item \textbf{SoRec}~\cite{ma2008sorec}: it integrates the social network between users and the user-item interaction matrix in the recommendation process, based on probabilistic matrix factorization.
\item \textbf{SoReg}~\cite{ma2011recommender}: it uses the user relations as the social regularization terms to constrain the matrix factorization objective.
\item \textbf{TrustMF}~\cite{yang2016social}: this approach fuses users' interactive behavior and trust relationships to conduct recommendations, by utilizing the matrix factorization to learn user embeddings in terms of their trust relationships.
\end{itemize}

\noindent \textbf{Graph Neural Network Collaborative Filtering Models}. We further compare \emph{\model} with two state-of-the-art recommendation models which augment the collaborative filtering architecture with graph-based neural techniques.
\begin{itemize}[leftmargin=*]
\item \textbf{STAR-GCN}~\cite{zhang2019star}: it introduces a stack of graph convolutional encoder-decoder to learn latent factors between users and items, with the reconstruction of masked embeddings.
\item \textbf{NGCF+SN}~\cite{wang2019neural}: NGCF is a state-of-the-art graph neural network-augmented collaborative filtering model under a message passing architecture. In order to incorporate the social network information into NGCF, we perform the embedding propagation on the integrative user-item and user-user relation graph, with the utilization of graph convolutional network to capture the high-order connectivity.

\end{itemize}

\noindent \textbf{Attentive Social Recommender Systems}. We further compare \emph{\model} with another line of social recommendation models which utilizes attention mechanism to encode the latent relationships between users and items.
\begin{itemize}[leftmargin=*]
\item \textbf{SAMN}~\cite{chen2019social}: it proposes a two-phase attention framework to capture relationships between users and identify the underlying informative signals from user's neighbors.
\item \textbf{EATNN}~\cite{chen2019efficient}: this approach is built upon an attention-based transfer neural network to adaptively learn the interplay relationships between the social and item domain.
\end{itemize}


\noindent \textbf{Social Recommendation with Graph Neural Networks}. Finally, we compare \emph{\model} with graph neural network-based social-aware recommendation framework.
\begin{itemize}[leftmargin=*]
\item \textbf{GraphRec}~\cite{fan2019graph}: it proposes a graph neural network model for social recommendation by aggregating social relations based on attention mechanism.
\item \textbf{DiffNet}~\cite{wu2019neural}: this recommendation method designs a layer-wise influence diffusion module to capture the influence propagation patterns between users in a recursive manner.
\end{itemize}

\begin{table*}[t]
\scriptsize
\vspace{-0.10in}
\caption{Performance comparison of all methods on three datasets in terms of \emph{RMSE} and \emph{MAE}.} 
\begin{center}
\setlength{\tabcolsep}{1.0mm}
\begin{tabular}{| c || c ||c|c|c| c | c | c | c | c | c | c | c | c || c|}
\hline
Data & Train  & Metrics & ~~PMF~~ & SocialMF & ~~SoReg~ & ~~SoRec~ & ~TrustMF~ & STA-GCN & NGCF+SN & ~~DiffNet~ & ~SAMN~ & EATNN & GraphRec & \emph{\model} \\
\hline
\multirow{4}{*}{Ciao} & \multirow{2}{*}{80\%}     & RMSE &  1.0664&	1.0657&	1.0782&	1.0526&	1.0518&	1.0295&	1.0306&	1.0369&	1.0543&	1.0313&	0.9687&	\textbf{0.9507}\\
\cline{3-15}
&                                                 & MAE  &  0.8281&	0.8321&	0.8593&	0.8135&	0.8113&	0.7687&	0.7781&	0.7723&	0.7890&	0.7667&	0.7382&	\textbf{0.7189}\\
\cline{2-15}
& \multirow{2}{*}{60\%}                           & RMSE &  1.0908&	1.0714&	1.0855&	1.0692&	1.0678&	1.0461&	1.0445&	1.0585&	1.0924&	1.0742&	0.9921&	\textbf{0.9624}\\
\cline{3-15}
&                                                 & MAE  &  0.8424&	0.8378&	0.8420&	0.8276&	0.8262&	0.7932&	0.7877&	0.7935&	0.8116&	0.7973&	0.7592&	\textbf{0.7318}\\
\hline
\multirow{4}{*}{Epinions} & \multirow{2}{*}{80\%} & RMSE & 1.1692&	1.1494&	1.1576&	1.1477&	1.1314&	1.0946&	1.1022&	1.1095&	1.1366&	1.1187&	1.0581&	\textbf{1.0326}\\
\cline{3-15} 
&                                                 & MAE  & 0.9187&	0.8730&	0.8797&	0.8732&	0.8642&	0.8582&	0.8650&	0.8438&	0.8671&	0.8545&	0.8074&	\textbf{0.7983}\\
\cline{2-15}
& \multirow{2}{*}{60\%}                           & RMSE &  1.1873&	1.1692&	1.1789&	1.1649&	1.1553&	1.1183&	1.1173&	1.1241&	1.1899&	1.1385&	1.0678&	\textbf{1.0411}\\
\cline{3-15}
&                                                 & MAE  &  0.9380&	0.8973&	0.9184&	0.8847&	0.8757&	0.8843&	0.8758&	0.8534&	0.8995&	0.8663&	0.8297&	\textbf{0.8081}\\
\cline{1-15} 
\multirow{4}{*}{Douban} & \multirow{2}{*}{80\%}   & RMSE &  0.7551&	0.7427&	0.7508&	0.7352&	0.7287&	0.7370&	0.7234&	0.7387&	0.7350&	0.7447&	0.7257&	\textbf{0.7141}\\
\cline{3-15}
&                                                 & MAE  &  0.5964&	0.5866&	0.5937&	0.5844&	0.5752&	0.5802&	0.5698&	0.5793&	0.5777&	0.5809&	0.5690&	\textbf{0.5645}\\
\cline{2-15}
& \multirow{2}{*}{60\%}                           & RMSE &  0.7674&	0.7589&	0.7624&	0.7459&	0.7377&	0.7531&	0.7305&	0.7524&	0.7483&	0.7619&	0.7348&	\textbf{0.7220}\\
\cline{3-15}
&                                                 & MAE  &  0.6063&	0.5983&	0.6074&	0.5924&	0.5826&	0.5909&	0.5760&	0.5885&	0.5879&	0.5934&	0.5787&	\textbf{0.5705}\\
\hline
\end{tabular}
\end{center}
\label{tab:result}
\end{table*}

\subsubsection{\bf Parameter Settings}
We implement our \emph{\model} with Pytorch and utilize Adam as the optimizer for model parameter inference. The hidden state dimensionality $d_r$ of our relation-aware graph neural module is tuned from the range of $[8,16,32,64,128]$. To achieve the trade-off between the social regularized representation and multi-relation encoding process~\cite{hjelm2018learning}, the embedding size $d_H$ in the mutual information-based social relation encoder is searched from the range of $[250, 500, 100, 1500, 2000]$. The batch size is chosen from $[1024, 2048, 4096, 8192]$ and the model optimization is performed with the learning rate of $1e^{-3}$. In our experiments, the early stopping is adopted to terminate the training process.


\subsection{Performance Comparison (RQ1)}
In table~\ref{tab:result}, we present the performance of all compared methods on three datasets, in terms of RMSE and MAE. In all cases, we can observe that \emph{\model} consistently outperforms different types of baselines by a significant margin. We attribute such improvement to the joint modeling of global social dependencies between users and multi-typed relations with respect to different user-item interactions. The performance is followed by GraphRec which models user-item relationships based on graph neural network. This verifies the utility of performing propagation information across users and items under a graph-structured learning framework. However, GraphRec fails to capture global social context when modeling social dependency-aware user's preference.

Among various baselines, we can observe that: by incorporating the social signals into the state-of-the-art neural graph collaborative filtering architecture (\ie, NGCF+SN), under a message passing framework with the relation heterogeneity and high-order connectivity over user-item graph, it could achieve competitive performance as compared to some deep social recommeder systems (\eg, SAMN and EATNN). This observation therefore points to the positive effect of modeling user-user and user-item graph-structured collaborative relations in the embedding function. The performance gap between attentive recommendation methods and graph neural network enhanced models also sheds light on the limitation of aggregating cross-domain dependencies with a weighted summation scheme. The potential reason lies in the failure to consider the high-level insights due to the hierarchical inter-dependencies across users and items.

\subsection{Model Ablation and Effectiveness Analyses (RQ2)}
To investigate the component-wise effect in our joint learning \emph{\model} framework, we consider different model variant settings from three perspectives and analyze their effects:

\begin{figure}
    \centering
    \vspace{-0.1 in}
    \subfigure[][Ciao-RMSE]{
        \centering
        \includegraphics[width=0.27\columnwidth]{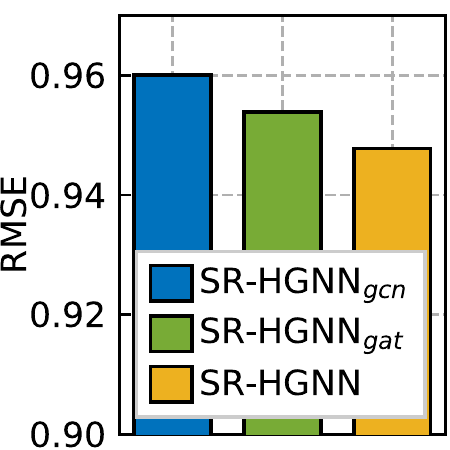}
        \label{fig:ablation_dgi_ciao}
        }
    \subfigure[][Epinions-RMSE]{
        \centering
        \includegraphics[width=0.27\columnwidth]{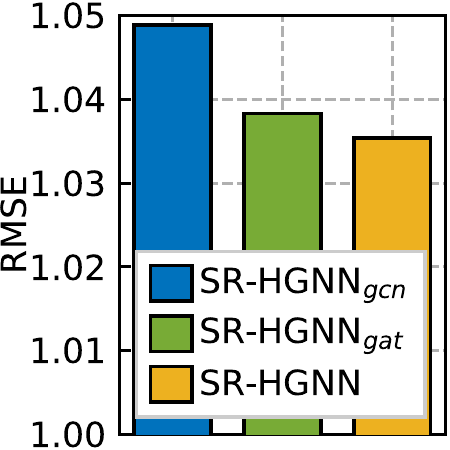}
        \label{fig:ablation_dgi_epinions}
        }
    \subfigure[][Douban-RMSE]{
        \centering
        \includegraphics[width=0.27\columnwidth]{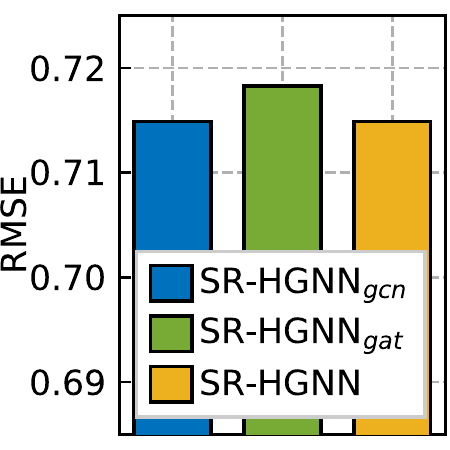}
        \label{fig:ablation_dgi_douban}
        }
    \subfigure[][Ciao-MAE]{
        \centering
        \includegraphics[width=0.27\columnwidth]{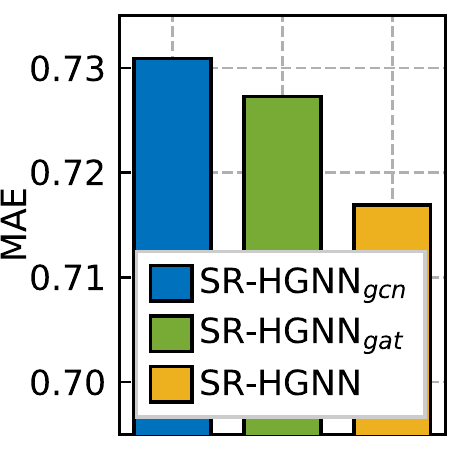}
        \label{fig:ablation_dgi_douban}
        }
    \subfigure[][Epinions-MAE]{
        \centering
        \includegraphics[width=0.27\columnwidth]{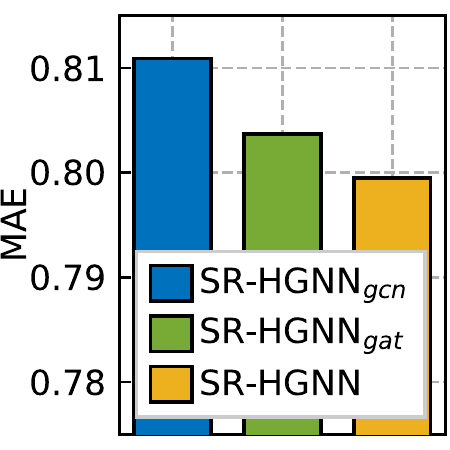}
        \label{fig:ablation_dgi_douban}
        }
    \subfigure[][Douban-MAE]{
        \centering
        \includegraphics[width=0.27\columnwidth]{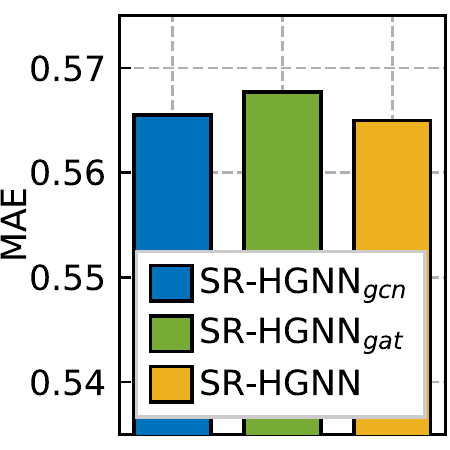}
        \label{fig:ablation_dgi_douban}
        }
    \caption{Ablation study on our mutual information-based graph neural module for social dependencies learning in terms of RMSE and MAE.}
    \label{fig:ablation_dgi}
\end{figure}

\subsubsection{\bf Global Social Relation Encoder}
To evaluate the effectiveness of our mutual information-based graph neural module in capturing global social dependencies, we first replace our social relation encoder with two representative graph neural network architectures: graph convolutional network~\cite{van2018graph} \emph{\model}$_{gcn}$ and graph attention network~\cite{wang2019heterogeneous} \emph{\model}$_{gat}$. The results are presented in Figure~\ref{fig:ablation_dgi}. It is clear to see that: while GCN and GAT have obtained promising results in fusing feature information between dependent users, our \emph{\model} could further boost the model accuracy through maximizing the mutual information between local and global representations of user dependence.


\begin{table}
	\footnotesize
	\vspace{-0.1in}
	\caption{Ablation test on the impact of the reconstruction-based context incorporation in terms of RMSE and MAE.}
	\vspace{-0.05in}
	\label{tab:ablation_recon}
	\centering
	\begin{tabular}{ccccc}
		\toprule
		Model&Metric&Ciao&Epinions&Douban\\
		\midrule
		\multirow{2}{*}{\emph{\model}$_{w/o-rec}$}&RMSE&0.9506&1.0378&0.7162\\
		&MAE&0.7203&0.8012&0.5668\\
		\hline
		\multirow{2}{*}{\emph{\model}$_{w-rec}$}&RMSE&0.9478&1.0354&0.7149\\
		&MAE&0.7169&0.7995&0.5650\\
		\bottomrule
	\end{tabular}
	\vspace{-0.2in}
\end{table}
\subsubsection{\bf Reconstruction-based Context Incorporation}
We further validate the impact of incorporating the reconstruction constrains (\ie, social-aware multi-relational information: reconstruction constrains $\mathcal{L}_{r}$ and $\mathcal{L}_{s}$) into our embedding learning process of users and items. Particularly, we generate the variant \emph{\model}$_{w/o-rec}$ without the context reconstruction component. The results in Table~\ref{tab:ablation_recon} show the benefit of the designed reconstructed graph neural module which endows \emph{\model} with the capability of characterizing overall cross-domain relational knowledge.

\begin{figure}[h]
    \centering
    \vspace{-0.1 in}
    \subfigure[][Ciao-RMSE]{
        \centering
        \includegraphics[width=0.27\columnwidth]{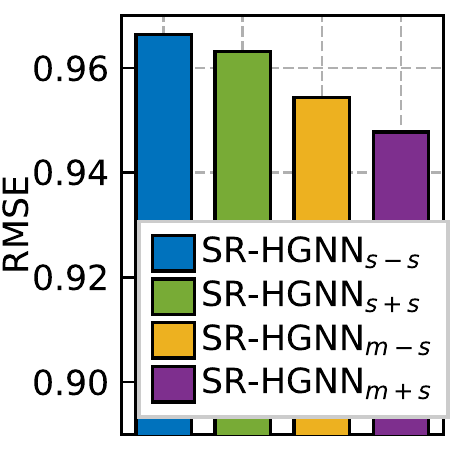}
        \label{fig:ablation_multi_ciao_rmse}
        }
    \subfigure[][Epinions-RMSE]{
        \centering
        \includegraphics[width=0.27\columnwidth]{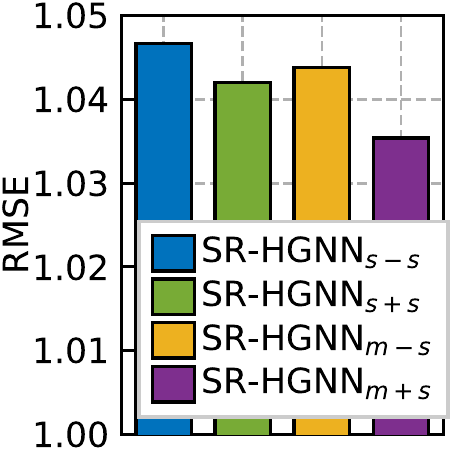}
        \label{fig:ablation_multi_epinions_rmse}
        }
    \subfigure[][Douban-RMSE]{
        \centering
        \includegraphics[width=0.27\columnwidth]{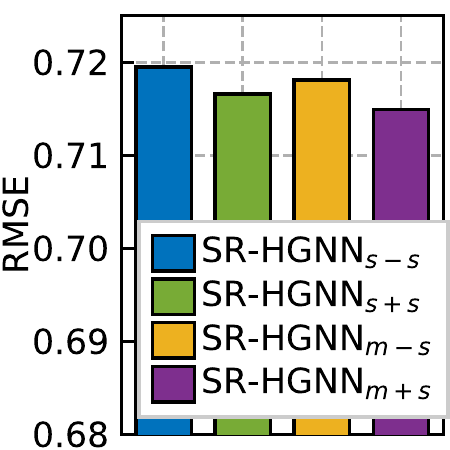}
        \label{fig:ablation_multi_douban_rmse}
        }
    \subfigure[][Ciao-MAE]{
        \centering
        \includegraphics[width=0.27\columnwidth]{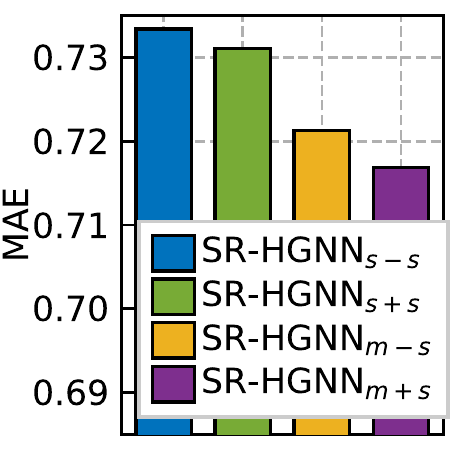}
        \label{fig:ablation_multi_ciao_mae}
        }
    \subfigure[][Epinions-MAE]{
        \centering
        \includegraphics[width=0.27\columnwidth]{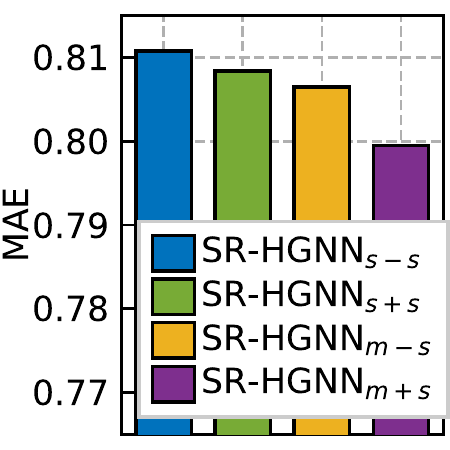}
        \label{fig:ablation_multi_epinions_mae}
        }
    \subfigure[][Douban-MAE]{
        \centering
        \includegraphics[width=0.27\columnwidth]{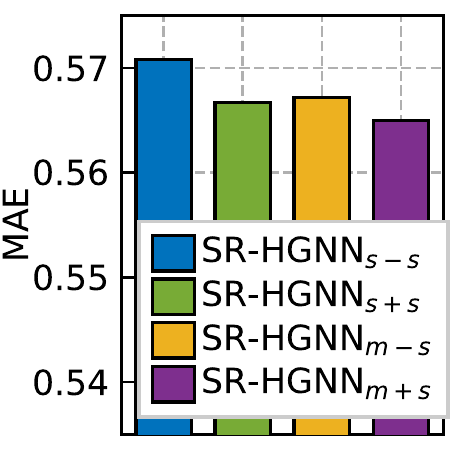}
        \label{fig:ablation_multi_douban_mae}
        }
    \caption{Ablation study on the effectiveness of the multi-typed user-item interactions and users' social information in terms of RMSE and MAE.}
    \label{fig:ablation_multi}
    \vspace{-0.05in}
\end{figure}

\subsubsection{\bf Multi-Typed Interactive Relation Learning}
Finally, we evaluate the influence of two key factors: i) multi-typed user-item interactions; and ii) social information among users. Accordingly, we generate four variants corresponding to these two dimensions: single-type interaction modeling with or without social information--\emph{\model}$_{s+s}$, \emph{\model}$_{s-s}$; multi-type interaction modeling with or without social information--\emph{\model}$_{m+s}$, \emph{\model}$_{m-s}$. From evaluation results in Figure~\ref{fig:ablation_multi}, we summarize two key observations:
\begin{itemize}[leftmargin=*]
\item We can first notice that the positive effect of social information in improving the recommendation performance.
\item The integration of multi-typed user-item relational structures with the recommendation framework could augment the learning process of complex user's preference.
\end{itemize}

\subsection{Performance \wrt\ Interaction Sparsity Levels (RQ3)}
\begin{figure*}
    \centering
    \vspace{-0.1 in}
    \subfigure[][Epinions]{
        \centering
        \includegraphics[width=0.29\textwidth]{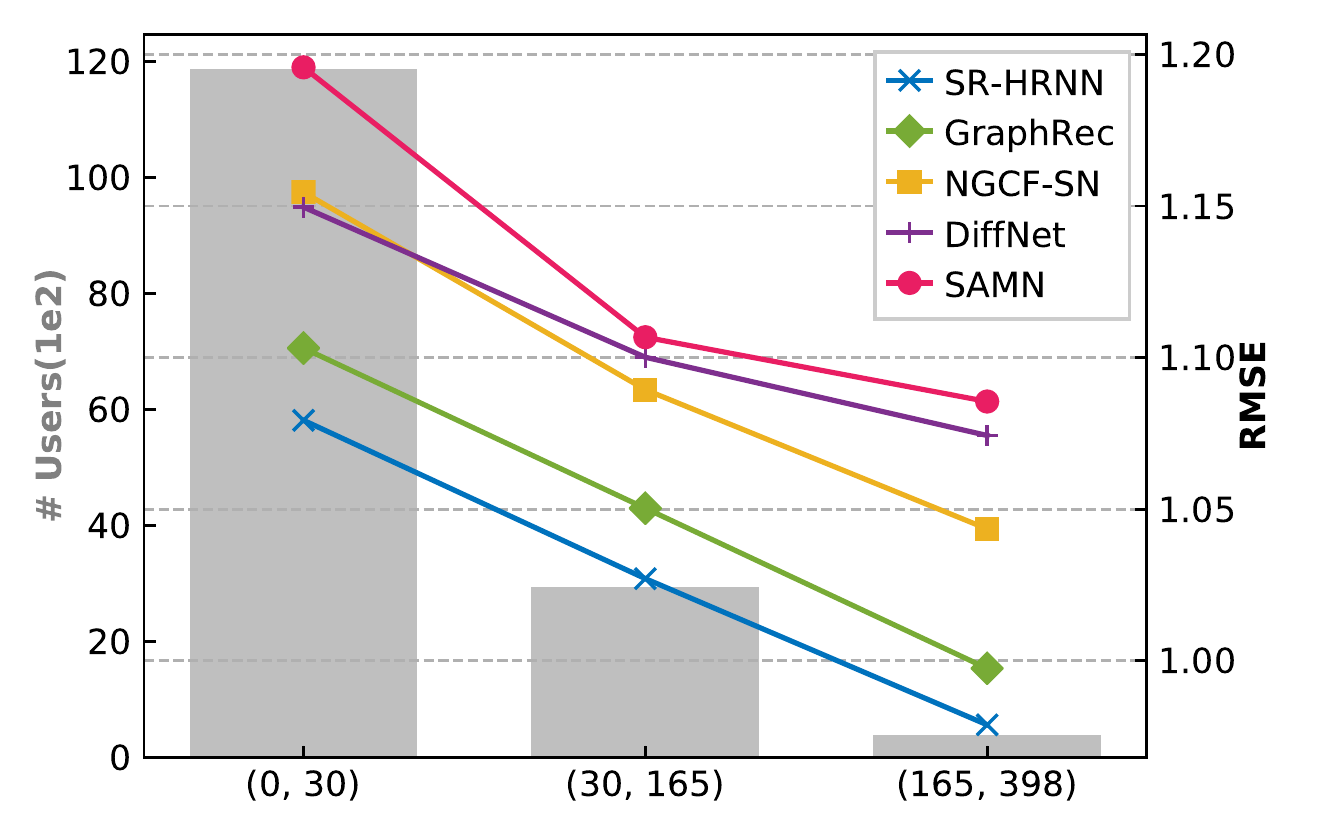}
        \label{fig:ML_1M_sparsity}
        }
    \subfigure[][Ciao]{
        \centering
        \includegraphics[width=0.29\textwidth]{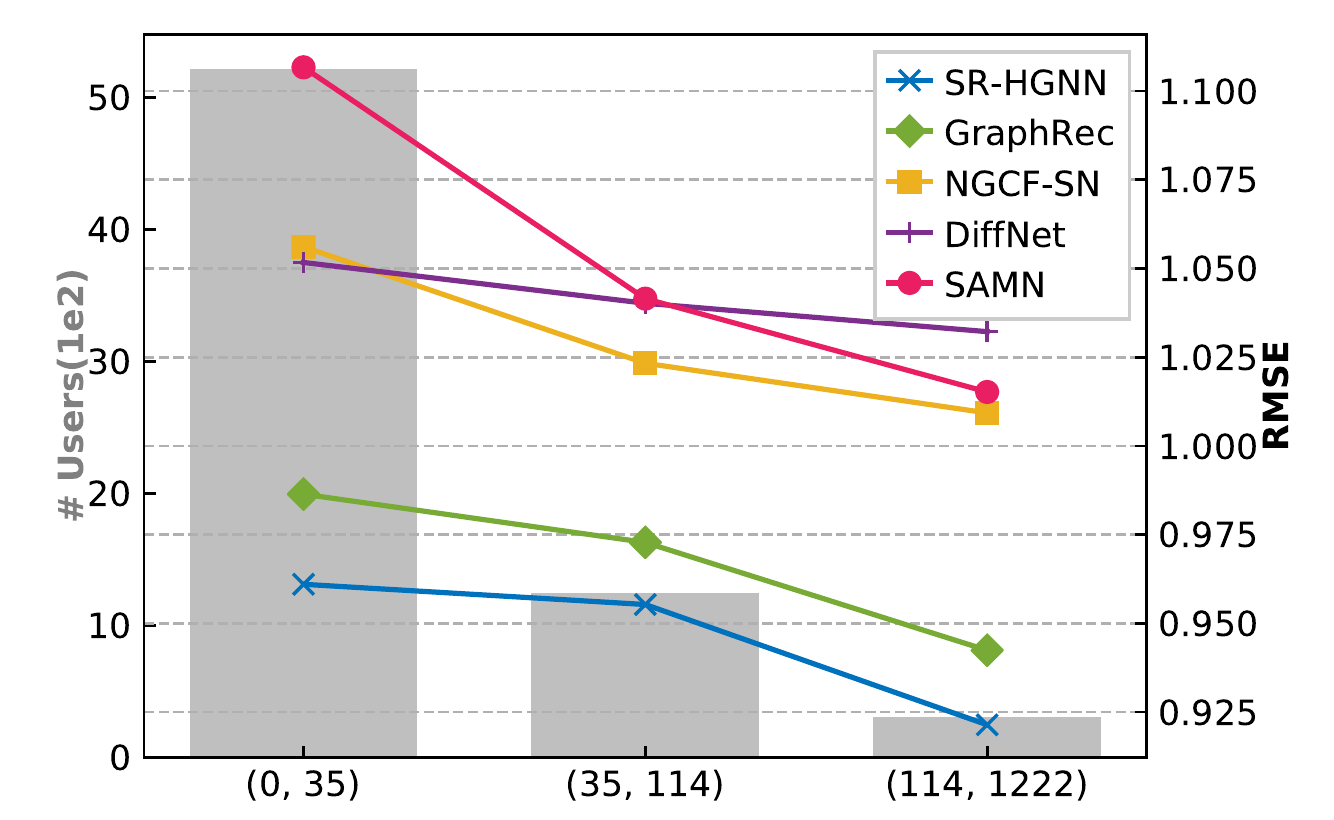}
        \label{fig:ML_10M_sparsity}
        }
    \subfigure[][Douban]{
        \centering
        \includegraphics[width=0.29\textwidth]{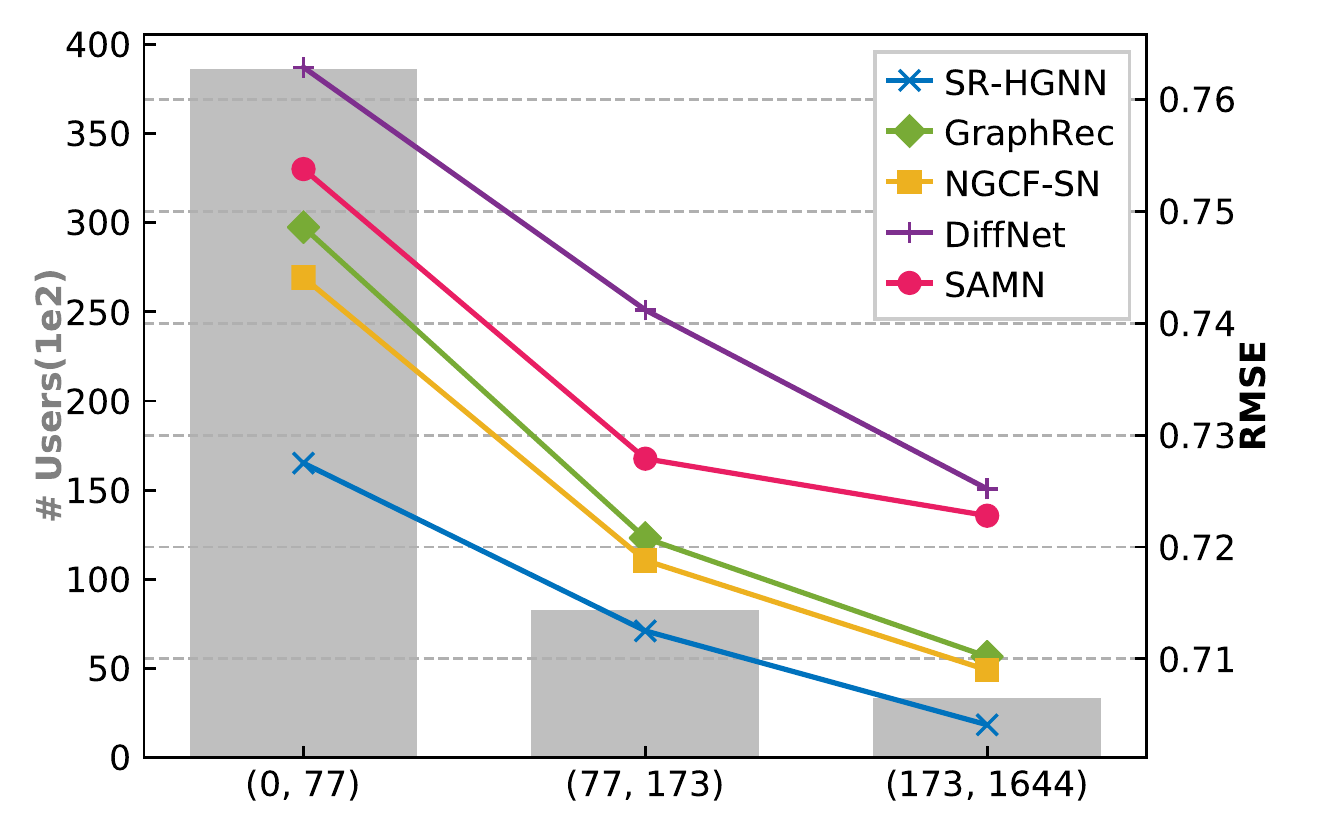}
        \label{fig:netflix_sparsity}
        }
    \vspace{-0.1 in}
    \caption{Performance comparison \wrt\ interaction sparsity levels, where background bars denote the number of users that falls into the specific interaction sparsity level, and the corresponding performance is represented by the lines.}
    \vspace{-0.1 in}
    \label{fig:sparsity}
\end{figure*}

We perform experiments to investigate the representation ability of our \emph{\model} in handling inactive users that interact with a limited number of items. As shown in Figure~\ref{fig:sparsity}, the recommendation performance is evaluated with respect to different interaction data sparsity levels. Specifically, we split target user instances into three groups with the increasing sparsity level, and keeping the number of interactions within each group to be equal. From the comparison results with several representative baselines, we can observe that \emph{\model} consistently outperforms competitive methods with different data sparsity levels. Moreover, larger performance gain can also be achieved by \emph{\model} in forecasting preference of sparse users on Douban dataset. These observations demonstrate that \emph{\model} is capable of effectively modeling relations from social and behavioral context modalities, to alleviate the data scarcity issue.

\subsection{Hyperparameter Study (RQ4)}
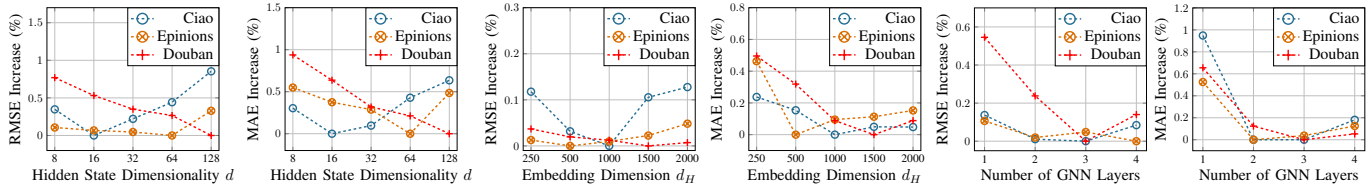
\begin{figure*}
    \centering
    \begin{adjustbox}{max width=1.0\linewidth}
    \input{./figures/parameters}
    \end{adjustbox}
    \vspace{-0.25in}
    \caption{Hyper-parameter study in terms of \textit{RMSE} and \textit{MAE}}
    \vspace{-0.1in}
    \label{fig:hyperparam}
\end{figure*}

We investigate the influence of hyperparameters in our \emph{\model} framework. To integrate results on different datasets with different performance scales into the same figure, we set y-axis as the performance variation ratio compared to the best performance. Figure~\ref{fig:hyperparam} shows the evaluation results. We summarize the key observations as follows:\\\vspace{-0.1in}

\noindent \textbf{Effect of dimensionality}. We separately evaluate the effects of $d$ and $d_H$, which corresponds to the hidden state dimensionality of our relation-aware reconstructed graph neural module and global social relation encoder. With the consideration of different dimension scales and concatenate operations, $d$ and $d_H$ is tuned with different embedding size ranges. We can observe that a larger value of hidden state dimensionality does not necessarily lead to better performance, due to the overfitting issue. We set $d$ and $d_H$ as $(16, 64, 128)$ and $(1000, 500, 1500)$ corresponding to Ciao, Epinions, Douban data, to achieve the best performance.\\\vspace{-0.1in}

\noindent \textbf{Effect of graph neural network depth}.
Increasing the depth $L$ of our relation-aware reconstructed graph neural module could improve the recommendation results. \emph{\model} with 2 and 3 embedding propagation layers obtain better performance as compared to the model which considers first-order relational structure only. We attribute such improvement to the high-order non-linearities brought by stacking more propagation layers. Additionally, the slight performance degradation can be noticed as $L$ increases since the deeper graph neural network tend to overfit.

\subsection{Model Efficiency Study (RQ5)}
\begin{figure}
    \centering
    \vspace{-0.1in}
    \subfigure[][Epinions]{
        \centering
        \includegraphics[width=0.225\textwidth]{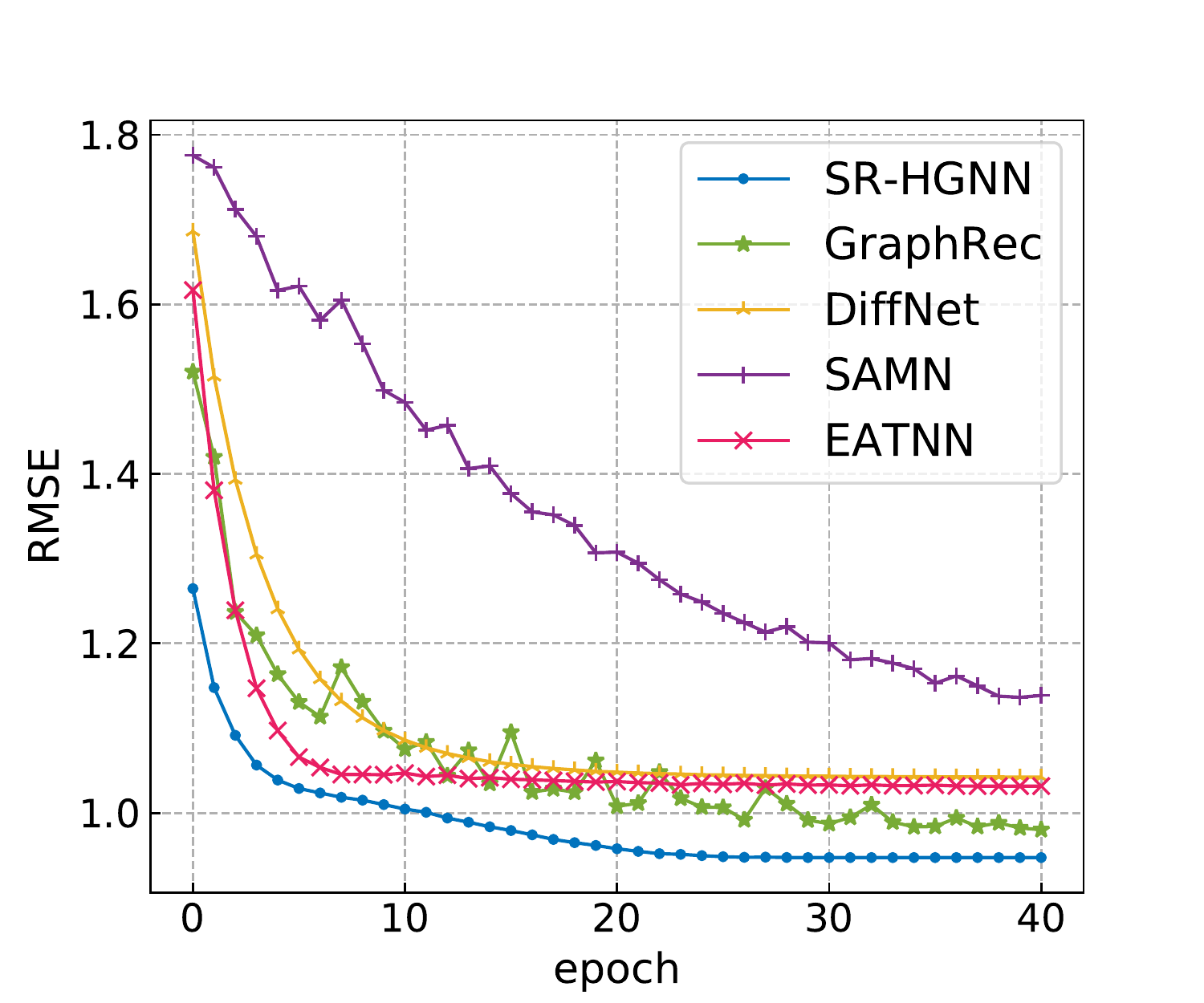}
        \label{fig:ML_1M_sparsity}
        }
    \subfigure[][Ciao]{
        \centering
        \includegraphics[width=0.225\textwidth]{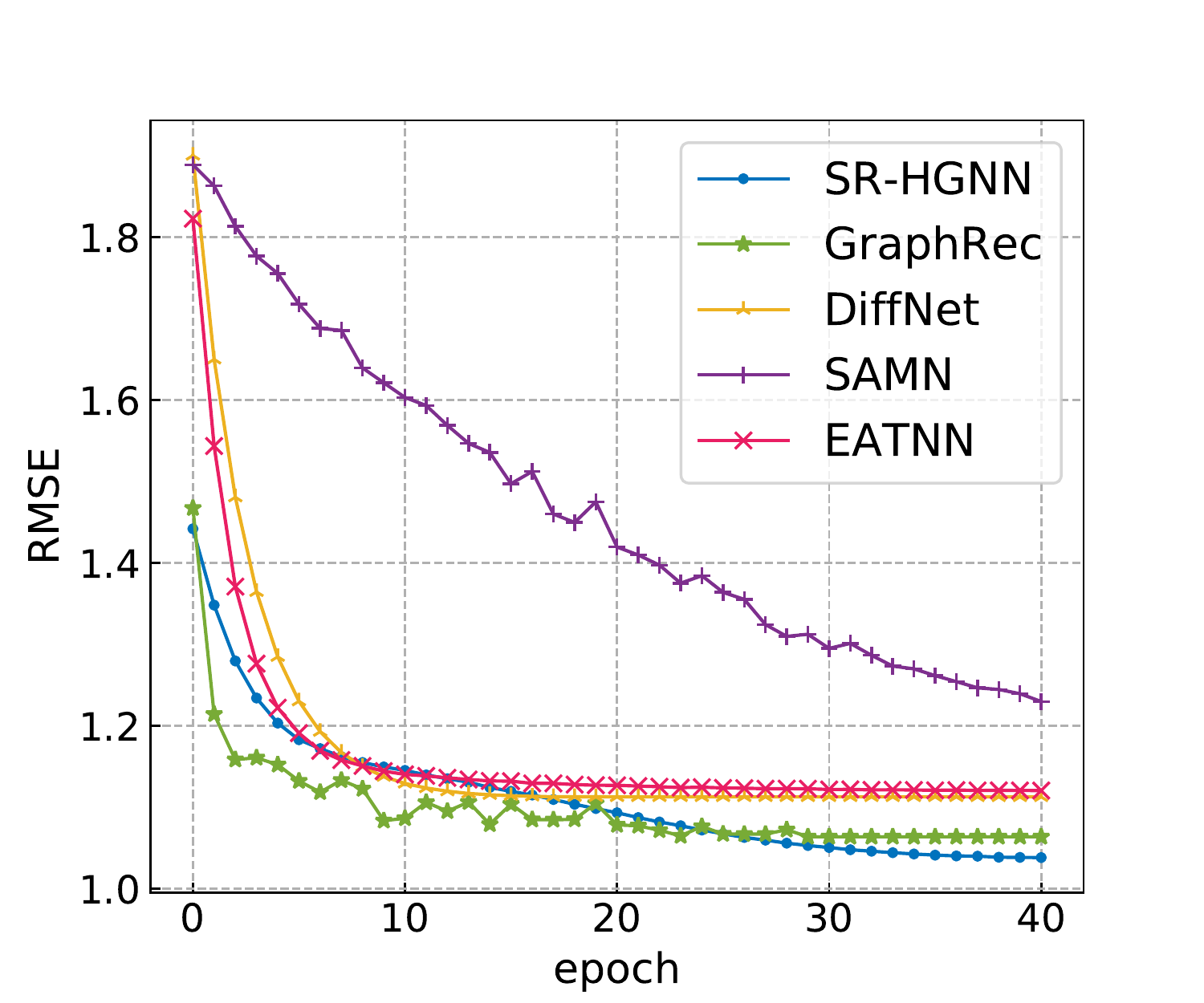}
        \label{fig:ML_10M_sparsity}
        }
    \caption{Convergence Study of \emph{\model} Framework.}
    \label{fig:convergence}
\end{figure}

We finally investigate the model efficiency of our \textit{\model}. Table~\ref{tab:time} presents the computational cost of training (with each individual epoch) for \textit{\model} and several deep neural network-based baselines on three different datasets. In our experiments, most compared baselines are evaluated using their released source codes. We can observe that \emph{\model} could achieve competitive efficiency as compared to most neural network-based social recommder systems. We also examine the convergence property of the \emph{\model}. Figure~\ref{fig:convergence} shows that the prediction accuracy as a function of the number of epochs. We could notice that \emph{\model} converges much smoother and faster than other state-of-the-art baselines in most cases, and performance is improved with more iterations. This observation also suggests the good efficiency of our social recommender system with hierarchical graph neural network.

\begin{table}
	\small
	\vspace{-0.05in}
	\caption{Model scalability study with running time (seconds).}
	\label{tab:run_time}
	\centering
	\begin{tabular}{cccc}
		\toprule
		Model & Ciao & Epinions & Douban \\
		\midrule
		NGCF+SN &3&12&63 \\
		DiffNet&2&3&10\\
		SAMN&3&6&20\\
		EATNN&1&2&5\\
		GraphRec&112&380&3600\\
        \hline
		\model&5&15&78\\
		\hline
	\end{tabular}
	\label{tab:time}
\end{table}

%% file: figures/parameters.tex


\begin{tikzpicture}
\begin{axis}[
    xlabel={Hidden State Dimensionality $d$},
    ylabel={RMSE Increase (\%)},
    xmin=0.8,xmax=5.2,
    ymin=-0.2,ymax=1.7,
    xtick={1,2,3,4,5},
    xticklabels={8,16,32,64,128},
    legend columns=1,
    legend cell align=right,
    grid=both,
    every axis plot/.append style={ultra thick},
    every tick label/.append style={scale=1.3},
    label style={scale=1.8},
    legend style={
        nodes={scale=1.5, transform shape},
        legend image post style={scale=1.5},
        },
    legend style={at={(1,1)},anchor=north east},
    every axis plot post/.append style={
        every mark/.append style={scale=2}
    }
]

\addplot[color={rgb:blue,4;green,2;yellow,1}, mark=o, dashed, mark options={solid}]
table[x=para, y=ciao_rmse] {latFactor.txt};
\addplot[color={rgb:red,4;green,1;yellow,2}, mark=otimes, dashed, mark options={solid}]
table[x=para, y=ep_rmse] {latFactor.txt};
\addplot[color={rgb:red,4}, mark=+, dashed, mark options={solid}]
table[x=para, y=douban_rmse] {latFactor.txt};
\legend{\large Ciao, \large Epinions, \large Douban};
\end{axis}
\end{tikzpicture}



\begin{tikzpicture}
\begin{axis}[
    xlabel={Hidden State Dimensionality $d$},
    ylabel={MAE Increase (\%)},
    ymin=-0.2,ymax=1.5,
    xmin=0.8,xmax=5.2,
    xtick={1,2,3,4,5},
    xticklabels={8,16,32,64,128},
    legend columns=1,
    legend cell align=right,
    grid=both,
    every axis plot/.append style={ultra thick},
    every tick label/.append style={scale=1.3},
    label style={scale=1.8},
    legend style={
        nodes={scale=1.5, transform shape},
        legend image post style={scale=1.5},
        },
    legend style={at={(1,1)},anchor=north east},
    every axis plot post/.append style={
        every mark/.append style={scale=2}
    }
]
\addplot[color={rgb:blue,4;green,2;yellow,1}, mark=o, dashed, mark options={solid}]
table[x=para, y=ciao_mae] {latFactor.txt};
\addplot[color={rgb:red,4;green,1;yellow,2}, mark=otimes, dashed, mark options={solid}]
table[x=para, y=ep_mae] {latFactor.txt};
\addplot[color={rgb:red,4}, mark=+, dashed, mark options={solid}]
table[x=para, y=douban_mae] {latFactor.txt};
\legend{\large Ciao, \large Epinions, \large Douban};

\end{axis}
\end{tikzpicture}


\begin{tikzpicture}
\begin{axis}[
    xlabel={Embedding Dimension $d_H$},
    ylabel={RMSE Increase (\%)},
    xmin=0.8,xmax=5.2,
    ymin=-0.01,ymax=0.3,
    xtick={1,2,3,4,5},
    xticklabels={250,500,1000,1500,2000},
    legend columns=1,
    legend cell align=right,
    grid=both,
    every axis plot/.append style={ultra thick},
    every tick label/.append style={scale=1.3},
    label style={scale=1.8},
    legend style={
        nodes={scale=1.5, transform shape},
        legend image post style={scale=1.5},
        },
    legend style={at={(1,1)},anchor=north east},
    every axis plot post/.append style={
        every mark/.append style={scale=2}
    }
]

\addplot[color={rgb:blue,4;green,2;yellow,1}, mark=o, dashed, mark options={solid}]
table[x=para, y=ciao_rmse] {dgi_latFactor.txt};
\addplot[color={rgb:red,4;green,1;yellow,2}, mark=otimes, dashed, mark options={solid}]
table[x=para, y=ep_rmse] {dgi_latFactor.txt};
\addplot[color={rgb:red,4}, mark=+, dashed, mark options={solid}]
table[x=para, y=douban_rmse] {dgi_latFactor.txt};
\legend{\large Ciao, \large Epinions, \large Douban};

\end{axis}
\end{tikzpicture}

\begin{tikzpicture}
\begin{axis}[
    xlabel={Embedding Dimension $d_H$},
    ylabel={MAE Increase (\%)},
    ymin=-0.1,ymax=0.8,
    xmin=0.8,xmax=5.2,
    xtick={1,2,3,4,5},
    xticklabels={250,500,1000,1500,2000},
    legend columns=1,
    legend cell align=right,
    grid=both,
    every axis plot/.append style={ultra thick},
    every tick label/.append style={scale=1.3},
    label style={scale=1.8},
    legend style={
        nodes={scale=1.5, transform shape},
        legend image post style={scale=1.5},
        },
    legend style={at={(1,1)},anchor=north east},
    every axis plot post/.append style={
        every mark/.append style={scale=2}
    }
]
]

\addplot[color={rgb:blue,4;green,2;yellow,1}, mark=o, dashed, mark options={solid}]
table[x=para, y=ciao_mae] {dgi_latFactor.txt};
\addplot[color={rgb:red,4;green,1;yellow,2}, mark=otimes, dashed, mark options={solid}]
table[x=para, y=ep_mae] {dgi_latFactor.txt};
\addplot[color={rgb:red,4}, mark=+, dashed, mark options={solid}]
table[x=para, y=douban_mae] {dgi_latFactor.txt};
\legend{\large Ciao, \large Epinions, \large Douban};
\end{axis}
\end{tikzpicture}

\begin{tikzpicture}
\begin{axis}[
    xlabel={Number of GNN Layers},
    ylabel={RMSE Increase (\%)},
    ymin=-0.05,ymax=0.7,
    xmin=0.8,xmax=4.2,
    xtick={1,2,3,4},
    xticklabels={1,2,3,4},
    legend columns=1,
    legend cell align=right,
    grid=both,
    every axis plot/.append style={ultra thick},
    every tick label/.append style={scale=1.3},
    label style={scale=1.8},
    legend style={
        nodes={scale=1.5, transform shape},
        legend image post style={scale=1.5},
        },
    legend style={at={(1,1)},anchor=north east},
    every axis plot post/.append style={
        every mark/.append style={scale=2}
    }
]

\addplot[color={rgb:blue,4;green,2;yellow,1}, mark=o, dashed, mark options={solid}]
table[x=para, y=ciao_rmse] {gnn_layer.txt};
\addplot[color={rgb:red,4;green,1;yellow,2}, mark=otimes, dashed, mark options={solid}]
table[x=para, y=ep_rmse] {gnn_layer.txt};
\addplot[color={rgb:red,4}, mark=+, dashed, mark options={solid}]
table[x=para, y=douban_rmse] {gnn_layer.txt};
\legend{\large Ciao, \large Epinions, \large Douban};

\end{axis}
\end{tikzpicture}

\begin{tikzpicture}
\begin{axis}[
    xlabel={Number of GNN Layers},
    ylabel={MAE Increase (\%)},
    ymin=-0.1,ymax=1.2,
    xmin=0.8,xmax=4.2,
    xtick={1,2,3,4},
    xticklabels={1,2,3,4},
    legend columns=1,
    legend cell align=right,
    grid=both,
    every axis plot/.append style={ultra thick},
    every tick label/.append style={scale=1.3},
    label style={scale=1.8},
    legend style={
        nodes={scale=1.5, transform shape},
        legend image post style={scale=1.5},
        },
    legend style={at={(1,1)},anchor=north east},
    every axis plot post/.append style={
        every mark/.append style={scale=2}
    }
]

\addplot[color={rgb:blue,4;green,2;yellow,1}, mark=o, dashed, mark options={solid}]
table[x=para, y=ciao_mae] {gnn_layer.txt};
\addplot[color={rgb:red,4;green,1;yellow,2}, mark=otimes, dashed, mark options={solid}]
table[x=para, y=ep_mae] {gnn_layer.txt};
\addplot[color={rgb:red,4}, mark=+, dashed, mark options={solid}]
table[x=para, y=douban_mae] {gnn_layer.txt};
\legend{\large Ciao, \large Epinions, \large Douban};

\end{axis}
\end{tikzpicture}

%% file: relate.tex
\section{Related Work}
\label{sec:relate}
This section discuss the research work which is related to our studied problem from the following aspects.
\vspace{-0.05in}

\subsection{Deep Collaborative Filtering Techniques}
Deep neural networks bring powerful representation and generalization ability for collaborative filtering recommendation techniques~\cite{zhang2019quaternion}. For example, NeurMF~\cite{he2017neural} replaced the inner-product operation with Multilayer Perceptron to learn non-linear relations between user and item embeddings. Inspired by the recent developments of graph neural networks, NGCF~\cite{wang2019neural} and STAR-GCN~\cite{zhang2019star} proposed to perform embedding propagation in the user-item integration graph. In addition, graph embedding technique has been leveraged to unify collaborative filtering with attention mechanism for pairwise user-item relation fusion~\cite{wang2019unified}. However, these models cannot well take the social relational information into consideration. Considering user-item interactive behavior tend to be influenced by other relevant users, this work incorporates social relations for recommendation by jointly modeling of multiplex user-item interactions and user-user social dependencies.

\subsection{Social-aware Recommender Systems}
Social recommendation aims at modeling the social signals among users to improve the recommender systems, with the consideration that users' interactions over items can be affected their friends~\cite{liu2019discrete,wang2017learning}. Previous work has made significant progress in incorporating social relationships into the matrix factorization framework with various integration schemes. For example, Ma~\etal~\cite{ma2008sorec} performed the factor analysis based on the probabilistic matrix factorization. Yang~\etal~\cite{yang2016social} built a matrix factorization model on the trust network of users based on their opinion propagation relationships. With the utilization of cross-domain user relations, multiple relations across heterogeneous networks were explored in~\cite{jiang2012social}.

With the advancement of deep learning techniques in revolutionizing recommender systems, many deep neural network models have been developed to jointly map user-item interactions and social relations between users into a shared latent space~\cite{wang2019social,yu2020enhance}. Among various algorithms, attention mechanism has served as an effective tool for relation aggregation. In particular, attentive memory mechanism has been utilized to learn influence strength among users in the social recommendation framework~\cite{chen2019social,chen2020social,chen2019efficient}. Fan~\etal~\cite{fan2019deep} proposed a random walk-based hierarchical attention network to select most relevant information from user's social neighbors. Furthermore, motivated by the idea of graph neural network for aggregating feature information from node's neighbors in a network structure, several attempts aim at aggregating user relations with graph attention encoder~\cite{song2019session,fan2019graph}. However, most of existing deep social recommender systems perform user aggregation via modeling the local behavioral similarity between users, which lacks an effective encoding of the collaborative signals between users in a comprehensive global space. To fill this gap, this work aggregates global contextual signals by exploring high-order relationships among users.

\subsection{Graph Neural Network for Recommendation}

Inspired by the promising results of graph neural network in learning dependence of graph structured data~\cite{wang2019heterogeneous,zhang2019heterogeneous}, another research line seeks to capture the user-item relationships with graph neural networks~\cite{wang2019neural,zhang2019star,ying2018graph}. For example, Wang~\etal~\cite{wang2019neural} proposed a graph relation encoder via the message passing between users and items for collaborative filtering. A stacked graph convolutional recommendation network was proposed to learn the masked user and item embeddings under a encoder-decoder architecture~\cite{zhang2019star}. PinSage~\cite{ying2018graph} applied the graph convolutional network in the user embedding generation process. Different from these models, \model\ augments the graph-based recommendation with the exploration of global contextual signals, based on the mutual information learning between low-level individual representations and high-level graph structure embedding.

%% file: conclusion.tex
\section{Conclusion}
\label{sec:conclusion}

In this paper, we proposed \model, to generalize graph neural network into social-aware collaborative filtering architecture, for jointly incorporating global dependencies between users and relation-aware users' preference over different items. In \model, we design a mutual information-contextualized social relation encoder which is capable of capturing global social dependencies among users. Based on the insight of multi-typed user-item interactions, we endow the graph-structured collaborative relation modeling to exploit the cross-interactive behavior dependencies. Our experiments show that \model\ consistently outperforms state-of-the-art social recommender systems. Future work includes incorporating external textual information of items (\eg, users' reviews or items textual descriptions) into the social recommendation framework to encode richer semantic signals. In addition, another line of future work lies in applying the developed \model\ framework to other types of social-aware recommendation datasets.